\documentclass[11pt,amsmath,amssymb,floatfix,twoside,cite]{revtex4}
\usepackage{graphicx}

\newcommand{\beq}{\begin{equation}}
\newcommand{\eeq}{\end{equation}}
\newcommand{\beqn}{\begin{eqnarray}}
\newcommand{\eeqn}{\end{eqnarray}}
\newcommand{\bearr}{\begin{array}}
\newcommand{\enarr}{\end{array}}

\newcommand{\toref}[1]{\mbox{(\ref{#1})}}

\begin{document}
\parskip 2mm

\title{Directed Percolation with long-range interactions:
 modeling non-equilibrium wetting}

\author{F. Ginelli$^{1}$, H. Hinrichsen$^{1}$, R. Livi$^{2,3}$,
 D. Mukamel$^{4}$ and A. Politi$^{2,5}$}
\affiliation{
$^1$Institut f\"ur Theoretische Physik und Astrophysik, University of
W\"urzburg, D-97974 W\"urzburg, Germany \\
$^2$ Istituto Nazionale di Fisica della Materia, Unit\`a di Firenze, Italy\\
$^3$ Dipartimento di Fisica, Universit\`a di Firenze, Italy \\
$^4$ Department of Physics of Complex Systems, The Weizmann
Institute of Science, 76100 Rehovot, Israel\\
$^5$ Istituto Nazionale di Ottica Applicata, Firenze, Italy}

\centerline{\date{\today}}

\begin{abstract}
It is argued that some phase--transitions observed in models of
non-equilibrium wetting phenomena are related to contact processes
with long-range interactions. This is investigated by introducing
a model where the activation rate of a site at the edge of an
inactive island of length $\ell$ is $1+a\ell^{-\sigma}$.
Mean--field analysis and numerical simulations indicate that for
$\sigma>1$ the transition is continuous and belongs to the
universality class of directed percolation, while for
$0<\sigma<1$, the transition becomes first order. This criterion
is then applied to discuss critical properties of various models
of non--equilibrium wetting.
\end{abstract}


\maketitle

\section{Introduction}

Many recent theoretical studies have shown that the growth process of a solid
phase on a substrate can undergo a variety of non-equilibrium transitions. They
are analogous to equilibrium wetting phenomena in which liquid boundary-layers
exhibit critical behavior in the vicinity of the liquid-gas coexistence line.
Such growth processes can be effectively modelled by defining a suitable
evolution rule for the profile (interface), corresponding to the boundary of
the solid layer. Several growth models which do not obey detailed balance and
evolve towards stationary  non--equilibrium states have been studied in the
past. In many cases, by varying a control parameter, they exhibit a  transition
from a regime where the solid phase remains pinned to the substrate to a regime
where an unbounded growth sets in
\cite{Alon,Wetting1,Wetting2,Wetting3,GiadaMarsili,Santos,CandiaAlbano,Ginelli03}.
This depinning of an interface may be considered as non-equilibrium wetting, in
analogy with its equilibrium counterpart. In some models the character of the
depinning transition changes, depending on the dynamical rates which control
the interface evolution. Numerical studies in $1+1$ dimensions allowed to
identify both first- and second-order phase--transitions,  those of the latter
type falling into two universality classes: directed percolation (DP)
\cite{Hinrichsen00} and multiplicative noise (MN) \cite{Genovese99}. Since the
transition always occurs towards an {\it irreversibly} depinned phase, it is
quite natural to draw an analogy with dynamical processes characterized by an
absorbing state. \\
Recently, it has also been shown that such a depinning transition can be related 
to the {\it synchronization} phase transition in spatially extended chaotic
systems \cite{Baroni, Pikovsky}. In this latter framework, two different
replicae of the same dynamical system are coupled one to each other, 
either deterministically \cite{Pikovsky} or by the addition of the same
realization of a spatiotemporal stochastic noise.
Upon increasing the coupling parameter, the system undergoes a non-equilibrium
phase transition between an unsynchronized phase and a completely synchronized
one, characterized by a vanishing local difference between the two
(initially) different replica. Since two completely synchronized replica of
the same system are bound to remain identical one with respect 
to each other at all future
times, we can conclude that the completely synchronized phase is absorbing for
system dynamics, thus playng the role of the irreversibly depinned phase
in non-equilibrium wetting processes. Interestingly, both numerical analyses
of coupled map lattices systems \cite{Baroni, Pikovsky} 
and analytical arguments \cite{Pikovsky, GinelliDP}
predict the synchronization transition to exhibit a critical
behavior, which belongs either to the MN or to the DP universality class.

\begin{figure}[tcb]
\centerline{\includegraphics[height=15mm]{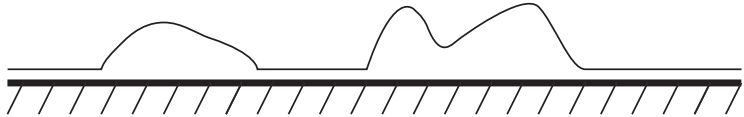}}
\caption{ \label{interface} \small Schematic configuration of an
interface bound to a substrate. Within the DP framework bound
sites may be considered as active while islands of depinned sites
are inactive.}
\end{figure}

A schematic typical configuration of an interface bound to a
substrate is depicted in Fig. \ref{interface}. It is composed of
detached domains separated by pinned segments. The dynamics of the
interface is such that each detached domain may either shrink or
expand from its edges; new domains may be created by the unbinding
process of bound sites, and two or more detached domains may merge
into a single larger one. In principle, a segment internal to a
detached domain may bind back to the substrate. However, in some
physical conditions such processes are virtually suppressed. This
occurs when the unbound interface moves on the average away from
the substrate, while it is held bound to the substrate by some
short range attractive interaction. In fact, the farther is a
segment from the edge of a domain, the larger is the height of the
inferface and, accordingly, the more unlikely the possibility to
bind back to the substrate. In this case, the dynamics of the
interface may very well be described by a contact process in which
the active sites correspond to those bound to the substrate. The
resulting depinning transition is thus expected to belong to the
DP universality class. The observation of DP depinning transition
in some models \cite{Alon,Ginelli03,Munoz} confirms the validity
of the above arguments. In some cases a possible crossover to a
first order transition has been suggested, when the attractive
interaction between the interface and the substrate is increased.
While in certain regions of the phase--diagram the existence of a
first order transition has been clearly demonstrated, in other
regions it has only been tentatively suggested, based on numerical
simulations of finite systems\cite{Wetting2,Munoz}. 
Due to the strict connection between wetting and complete synchronization 
non-equilibrium phase transitions, a more accurate understanding of this 
part of the wetting phase diagram is highly desirable not only from the theoretical
point of view, but also in view of possible experimental realizations
of the MN and/or the DP universality classes. In our opinion, in fact, the 
synchronization transition in spatially extended systems represents the most 
promising framework in which one can look for MN critical properties, if not for DP
ones, which seem to require a highly non linear local dynamics governing
a spatially extended system \cite{Baroni, GinelliDP}. It is worth stressing
that both these classes are still eluding a clear experimental evidence,
possibly due to the presence of quenched disorder in experimental realizations
and, in the case of DP phase transitions, to seemingly unavoidable
small fluctuations of the (supposedly) absorbing state \cite{Hinrichsen00}. 
Interesting candidates for experiments include the photosensitive
Belousov-Zhabotinsky (BZ) reaction, which is known to exhibit complex 
spatiotemporal dynamics \cite{BZ} and semiconductor lasers
with time-delayed optical feedback. In Ref. \cite{GiacomelliPoliti}
it has been shown that such delayed systems can be interpreted in terms
of a suitable spatiotemporal dynamics, where the effective system size 
is given by the ratio between the delay time and the typical fast timescale 
of the system. 
Notice that, in principle, the use of semiconductor lasers
with time--delayed feedback allows for obtaining rapidly
a large data set, which is basically free of quenched noise
effects. These features provide the possibility of a proper
statistical description of the synchronization transition.
Finally, the role played by small 
fluctuations near the completely synchronized phase and their exponential
suppression in the case of synchronization DP has been discussed in Ref. \cite{GinelliDP}.\\

In the present
paper we introduce a framework within which the crossover from DP
to first order transition may be examined. This framework is then
applied to two previously introduced models of non-equilibrium
wetting.

Preliminary simulations performed in the pinned phase close to a seemingly
first-order wetting transition have revealed that the activation rate at the
border of a depinned island depends on the island length. This suggests that
the dynamics of a fluctuating interface leads to an effective interaction
between the sites at the island--boundaries. If this interaction is long range,
it can affect the dynamics of large islands and, in principle, provide a
mechanism for a first order wetting transition. In this paper we consider a
generalization of the contact process where the activation rate of sites at the
boundary of an inactive domain decays algebraically with the domain length.
In particular, the activation rate for an island of length $\ell$ is assumed to
take the form $ \lambda (1 + a/\ell^\sigma)$, where $\lambda$, $\sigma$ and $a$
are positive constants. We find that, depending on the power $\sigma$, the model
exhibits either a continuous DP-like (for $\sigma>1$) or a first-order (for
$\sigma < 1$) transition. We then examine the possible emergence of such
effective long range interactions in wetting models. Numerical determination of
the power $\sigma$  of the effective interaction thus allows inferring the order
of the phase transition also in these models.

The paper is organized as follows: In Section 2 we introduce a
generalized contact process in $1+1$ dimensions that includes an
activation rate which decays algebraically with the size of
inactive domains. The mean--field solution of the model predicts a
first--order transition, when the interaction decays slowly enough
with the inactive domain size, while faster decay rates yield a DP
behavior. In Section 3 we present the results of detailed
numerical simulations, which are in a very good agreement with the
mean field predictions. The bridge between this generalized
contact process and wetting models is discussed in Section 4.
There we consider two models, which have been introduced
previously for studying non-equilibrium wetting. The first is a
solid-on-solid model~\cite{Wetting1,Wetting2,Wetting3}, whose
phase--diagram contains a line of wetting transitions. While the
first order nature of the transition has clearly been established
in a part of the phase--diagram, the nature of the transition in
another region proved to be more difficult to analyze. Our
numerical study seems to suggest that inside the latter region,
upon changing a suitable control parameter, the effective
activation rate at the boundary of inactive domains can exhibit
both a fast ($\sigma > 1$) and a slow ($\sigma < 1$) power law
decay. However further analysis based on scaling arguments shows
that the slow decay of the effective activation rate (and thus the
first order behavior) is only a finite-size effect, albeit
particularly robust. This result suggests that in this entire
region the transition is asymptotically continuous and of DP
nature. We then analyze a second wetting model
(single-step-with-wall)~\cite{Ginelli03}, where previous studies
have indicated a continuous DP-like wetting transition for strong
attractive interaction between the substrate and the interface. We
provide numerical evidence that the effective activation rate
governing the dynamics of inactive islands corresponds to the case
$\sigma > 1$. This is indeed consistent with the DP nature of the
transition. The main results are summarized in Section 5, which
contains also some remarks and comments on future perspectives.

\section{A contact process with long range interactions}

In this section we introduce a lattice model of a contact process in $1+1$
dimensions with long range interactions which shows a crossover between a
continuous DP and a first--order phase--transition. We consider a periodic
lattice of length $L$, where the state variable $S_i$ at site $i$ is either
``active'', $S_i=1$, or ``inactive'', $S_i=0$. The dynamics evolves by
random-sequential updates, i.e. at each time step a lattice site $i$ is chosen
at random. If the selected site is either active or next neighbor of an active
site, it is updated according to the following rules

\begin{eqnarray}
\label{Process}
1  &\rightarrow& 0  \hspace{0.5cm}\mbox{with rate }\hspace{0.5cm} 1 \\
0^{\ell}1 \hspace{0.25cm} [10^{\ell}] &\rightarrow& 0^{\ell -1}11 \hspace{0.25cm}
[110^{\ell -1}]
\hspace{0.5cm}\mbox{with rate }\hspace{0.5cm}
\bar{\lambda}(\ell)=\lambda(1+a/\ell^\sigma)\,,
\nonumber
\end{eqnarray}
where $0^{\ell}$ is a shorthand notation for an inactive island of
size $\ell$. Finally, as for usual contact processes, inactive
sites that are not adjacent to active ones cannot be activated,
thus guaranteeing that the inactive state is absorbing. The
constants $\lambda$ and $a$ are both non--negative: the case $a=0$
corresponds to the usual short range contact process, which
exhibits a DP transition at $\lambda_c =
1.64892(8)$~\cite{MarroDickman99}. For $a > 0$ the power law decay
with $\ell$ of the shrinking rate of inactive islands introduces
effective long-range interactions. Within the mean field
approximation one finds that the transition is continuous for
$\sigma > 1$ but becomes first--order for $0<\sigma<1$. To
demonstrate this point let $\rho(t)$ be the average density of
active sites at time $t$. In the thermodynamic limit, $L \to
\infty $, the mean field dynamics of $\rho$ reads
\begin{equation}
\frac{d \rho}{d t} =
-\rho + \lambda \rho^2 \sum_{\ell=1}^{\infty}
(1+\frac{a}{\ell^{\sigma}})(1-\rho)^\ell =  (\lambda-1)\rho -\lambda \rho^2 + \lambda a \rho^2 \sum_{\ell=1}^{\infty}
\frac{(1-\rho)^\ell}{\ell^{\sigma}} \,.
\label{MFsigma}
\end{equation}
For $\sigma > 1$ the sum on the r.h.s. of Eq.~\toref{MFsigma} is finite in the
limit $\rho \to 0$ and its contribution amounts to a renormalization of the
coefficient of the $\rho^2$ term. Accordingly, the mean field equation
describes a standard DP process with short range interaction, thus recovering,
for sufficiently small values of $a$, the continuous nature of the
phase--transition. For large $a$, the coefficient of the  $\rho^2$ term becomes
positive and the mean--field approximation predicts a first--order transition.
However, studies of similar models indicate that the mean--field prediction
of first--order phase--transitions associated with the change of
sign of the $\rho^2$ term is unreliable~\cite{HayeUnpublished}.

For $0<\sigma<1$,  the leading contribution arising from the sum in the
rightmost side of Eq.~\toref{MFsigma} can be captured by replacing it
with an integral over $d\ell$,
\begin{equation}
\partial_t\rho = (\lambda-1)\rho -\lambda \rho^2 + \lambda a \rho^2 \int_0^{\infty}
\frac{(1-\rho)^\ell}{\ell^{\sigma}} \,d\ell\,,
\label{MFsigma2}
\end{equation}
which, to leading order in $\rho$, reduces to
\begin{equation}
\partial_t\rho = (\lambda-1)\rho +
a \lambda \Gamma(1-\sigma)\rho^{1+\sigma}-\lambda \rho^2 \,.
\label{MFsigma3}
\end{equation}
Here $\Gamma(x)$ is the standard Gamma function. The leading nonlinear term in
this equation involves a non-integer power, as a consequence of the long range
nature of the interactions. Since its coefficient is positive,
Eq.~\toref{MFsigma3} cannot admit fixed point solutions for arbitrarily small
densities, so that the transition to the absorbing state is first order. This
result is expected to hold even for $\sigma =1$, where the leading singular
term in the equation is $-\rho^2 \ln \rho$.

\section{Numerical results}

The mean field calculation, that is expected to hold above the
upper critical dimension of directed percolation $d_c=4$,
indicates that for $\sigma>1$ the transition is second--order with
the critical exponents of DP, while for $\sigma\leq1$ it turns
into a first--order transition. The crossover from a continuous to
a discontinuous transition is therefore predicted to take place at
$\sigma_c=1$. In what follows we investigate the validity of this
prediction by simulating the generalized contact process defined
in Eq.~(\ref{Process}). In order to obtain independent checks, we
performed two different kinds of numerical analyses, i.e.
measuring the scaling properties of suitable observables starting
from i) a fully active state and ii) a single active site
(epidemic spreading). Although we do not expect the overall
scenario to depend on the parameter $a$ (see
Eq.~\toref{Process}~), the accuracy of the numerical simulations
actually does depend. For small $a$, discontinuities in the order
parameter are correspondingly small, while for large $a$, it is
necessary to consider very large lattice sizes to reach the
asymptotic--scaling regime. All the numerical results reported in
this paper have been obtained for $a=2$, which represents a good
compromise. We find $\sigma_c = 1.0 \pm 0.1$, which suggests that
the mean--field analysis is quite accurate even in $1+1$
dimensions. However, we cannot exclude the possibility that the
precise threshold value in such a low dimensional case slightly
deviates from $1$. This has already been observed in directed
percolation with long-range infections through Levy flights, where
a small deviation from the mean--field prediction was found for
the critical value of the control parameter (i.e., the exponent of
the Levy distribution)~\cite{JanssenEtAl99,HinrichsenHoward99}.

\begin{table}[tbc]
\begin{center}
\begin{tabular}{c|cccccc}
$\sigma$ & 0.5 & 0.8 & 0.9 & 1.1 & 1.2 & 1.5 \\
\hline
$\lambda^{f}_c$ & 1.2850(5)& & 1.3650(5) & 1.4091(1) & & 1.4708(3) \\
$\lambda^{l}_c$ & 1.2870(4)& 1.3395(3) & 1.3635(5) & 1.4093(3) & 1.4280(5)
                  & 1.4711(3)
\end{tabular}
\end{center}
\caption{\label{TABQ} Estimates of the critical points for $a=2$.
They have been obtained starting from a fully active initial state
($\lambda^{f}_c$) and by spreading analysis of localized initial
conditions ($\lambda^{l}_c$). Finite size corrections are responsible
of the small differences.}
\end{table}
%
%
\subsection{Analysis of the stationary active state}
We first discuss the evolution of Monte Carlo simulations that start from a
fully active state. In order to distinguish between first-order and
second-order transitions we determine the density of active sites and the
size-distribution of inactive islands. Let us first summarize the expected
results for each of the two quantities. Since the continuous transition should
be DP, the average density of active sites, $\rho(t)$, measured at criticality,
should decay as \beq \rho(t) \sim t^{-\theta} \quad. \eeq On the other hand,
off-criticality, in the active phase of an infinite system, $\rho(t)$ saturates
to a stationary value $\langle \rho \rangle_t$ (where $\langle \cdot \rangle_t$
denotes time average), which scales with the distance from the critical point
$\lambda_c$ as
\beq \langle \rho \rangle_t \sim |\lambda - \lambda_c|^{\beta}
\quad, \label{betascaling} \eeq
where $\beta=0.276486(8)$ \cite{Jensen99}. These two critical exponents are
connected by the so called temporal exponent $\nu_{\parallel}$, so that
$\theta=\beta/\nu_\parallel=0.159464(6)$~\cite{Jensen99}.\\
Moreover, in DP the size--distribution $P(\ell)$ of inactive islands decays
algebraically as
\beq
\label{elle} P(\ell) \sim \ell^{-(2-\beta/\nu_{\perp} )}
\eeq
with a cutoff at the spatial correlation length
$\xi_\perp \sim |\lambda - \lambda_c|^{-\nu_{\perp}}$, where
the exponent $2-\beta/\nu_\perp = -1.747 \ldots$ follows from
$\langle \ell \rangle \sim 1/ \rho \sim \xi_\perp^{\beta/\nu_\perp}$.
In fact, the average length of inactive islands \beq \langle \ell
\rangle = \int_0^{\infty} P(\ell) \ell\, d\ell \label{laverage}
\eeq diverges at criticality due to scale invariance, so that
$P(\ell) \sim \ell^{-\alpha}$ with $1 < \alpha < 2$ for $\ell
\lesssim \xi_{\perp}$. Since the average size of inactive
islands is proportional to the inverse of the density of active
sites, one has $1 / \rho \sim \langle \ell \rangle \sim
\xi_{\perp}^{\beta / \nu_{\perp}}$, which implies $\alpha = 2 -
\beta / \nu_{\perp} = 1.747\ldots $ \cite{foot1}.

\begin{figure}[tcb]
\vglue 4mm
\centerline{\includegraphics[height=60mm]{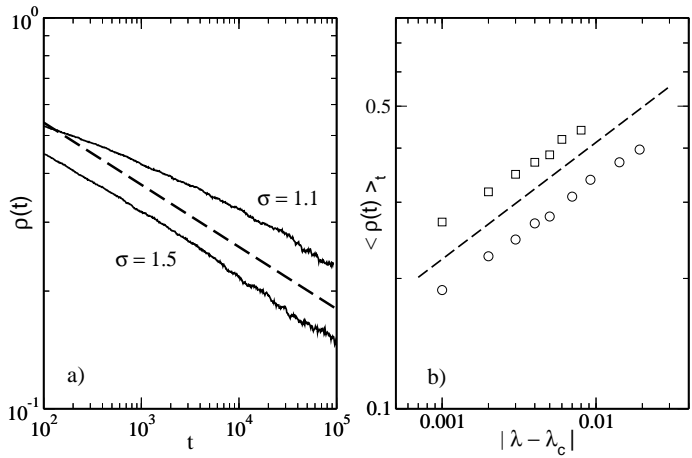}}
\caption{
\label{Figdelta} \small
Power law critical behavior of the generalized contact process
(\ref{Process})
for $\sigma = 1.1$ and $\sigma = 1.5$, compared to
the expected DP behavior (dashed lines):
a) decay in time of the density of active sites $\rho(t)$;
b) stationary density of active sites as a function of the distance from
criticality. Circles correspond to the case $\sigma = 1.5$, while squares
to $\sigma = 1.1$. Both graphs are plotted in a doubly logarithmic scale.
}
\end{figure}

Conversely, for a first--order phase--transition $\rho(t)$ is not expected to
exhibit any critical behavior associated with a diverging correlation length at
the transition point. Instead, the saturated order parameter $\langle \rho
\rangle_t$ exhibits a discontinuity at the transition. Since in this case the
active phase cannot display any coarsening properties, the average length
\toref{laverage} of inactive islands should be finite, i.e. $P(\ell)$ should decay
{\em faster} than $1/\ell^2$. As no hysteretic behavior can be observed in
non-equilibrium processes with an absorbing state, the distribution $P(\ell)$
turns out to be the most effective indicator of a first-order transition.

\begin{figure}[tbc]
\centerline{\includegraphics[height=60mm]{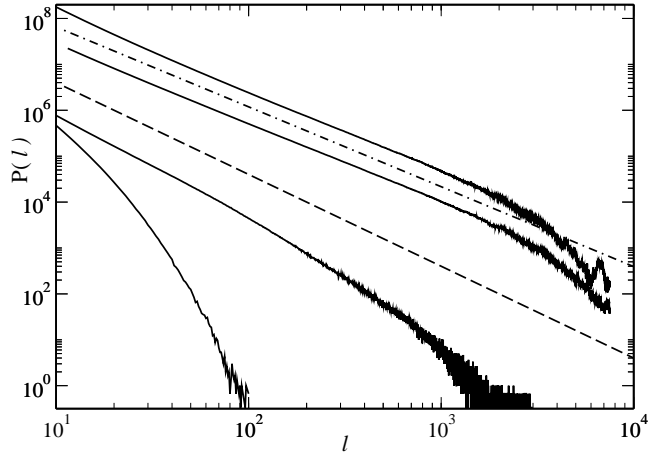}}
\caption{ \label{Figislands} \small Doubly logarithmic graph of
the (un-normalized) size distributions of inactive islands
$P(\ell)$ as a function of size $\ell$. From top to bottom the
solid lines corresponds to $\sigma=1.5$, $\sigma=1.1$,
$\sigma=0.9$ and $\sigma=0.5$. The dot-dashed line marks the
power-law decay expected in the case of a DP phase--transition. The
long dashed line decays as $1 / \ell^2$, discriminating in 1+1
dimensions between first order and continuous phase--transitions. }
\end{figure}

In order to reduce as much as possible finite--size effects, we
considered very large systems of size $L=2^{19}$ with periodic
boundary conditions (additionally we have also averaged over a few
different realizations). The best estimates of the critical point
$\lambda^{f}_c$, are reported in the first row of
Table~\ref{TABQ}. For $\sigma=1.1$ and $\sigma=1.5$ we identified
a critical scaling region (see Fig. \ref{Figdelta}), where both
the exponents $\delta$ and $\beta$ are in agreement with the best
numerical estimates for DP in 1+1 dimensions. Moreover, we
computed $P(\ell)$ in the active phase close to the critical point
by sampling spatial configurations at periodic time intervals and
counting all the inactive regions of size $\ell$. As shown in
Fig.~\ref{Figislands}, $P(\ell)$ is characterized by a power law
decay slower than $1/\ell^2$ (prior to the unavoidable exponential
cutoff due to finite--size effects), consistently with the
prediction for DP (see Eq.~\toref{elle}).

On the other hand, below $\sigma = 1$
a scaling region at the transition point could not be identified.
The saturated density of active sites $\langle \rho \rangle_t$
shows a finite discontinuity and $P(\ell)$ decays faster than $1/\ell^2$
~(see Fig. \ref{Figislands})~.
Accordingly, this analysis provides evidence that the transition
is discontinuous.

\subsection{Spreading from a single seed}
\label{spread}

A further verification of the mean--field analysis has been obtained by
simulating model \toref{Process}, starting from a single active site at the
origin~\cite{GrassbergerTorre79}. In this type of simulations, the relevant
variables are the survival probability $P_s(t)$, the number $N(t)$ of active
sites (averaged over all runs) and the mean square spreading $R^2(t)$ of the
active region. At the critical point of a phase--transition towards an
absorbing state, these quantities are known to scale as
\begin{equation}
\label{spreadEQ}
P_s(t) \sim t^{-\delta}\,, \quad N(t) \sim t^\eta\,, \quad R^2(t) \sim t^{2/z}\,.
\end{equation}
In the special case of a DP phase--transition, the exponents $\delta,\eta,z$
can be expressed in terms of the standard
exponents $\beta,\nu_\perp,\nu_\parallel$ as \cite{foot2}
\begin{equation}
\delta=\theta=\beta/{\nu_\parallel},\quad
\eta=(d\nu_\perp-2\beta)/\nu_\parallel,\quad
z=\nu_\parallel/\nu_\perp.
\label{Hyperscaling}
\end{equation}
\begin{figure}[tcb]
\centerline{\includegraphics[width=90mm]{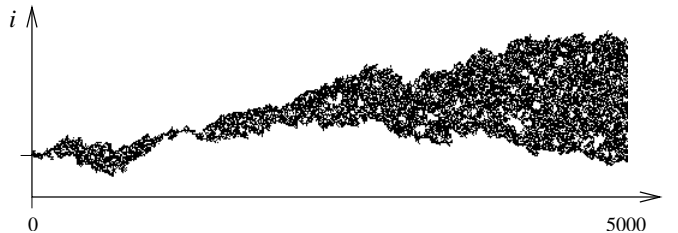}}
\caption{
\label{FIGDEMO} \small
Typical cluster grown from a single seed for $\sigma=0.5$ at criticality,
simulated up to 5000 Monte Carlo sweeps. Notice that large inactive islands
are suppressed and the growing cluster can be regarded as effectively compact.
}
\end{figure}
\begin{table}[tcb]
\begin{center}
\begin{tabular}{c|ccc}
&$\delta$ &$\eta$ & $z$\\
\hline
DP & 0.159464(6) & 0.313686(8) & 1.580745(10) \\
Glauber & 1/2 & 0 & 2
\end{tabular}
\end{center}
\caption{\label{TabSeed}
Exponents in seed simulations for DP and zero-temperature Glauber dynamics in
1+1 dimensions.}
\end{table}
\begin{figure}[tcb]
\includegraphics[width=125mm]{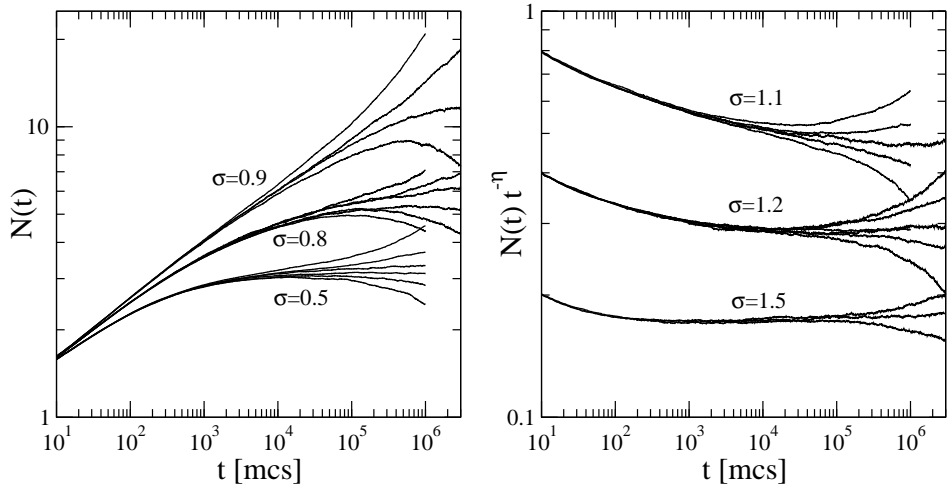}
\caption{
\label{FIGN} \small
Time evolution of the average number of active sites close to
the phase transition.
In the left panel the average number of active sites $N(t)$ is plotted as a
function of time
for $a=2$, $\sigma=0.5, 0.8, 0.9$ and for different
values of $\lambda$, close to $\lambda_c$. All curves indicate
that $N(t)$ converges to a constant after a transient time
which increases with $\sigma$.
In the right panel the time--rescaled number of active sites
$N(t) t^{-\eta}$ (with $\eta = 0.313686$) is showed
for $a=2$, $\sigma=1.1, 1.2, 1.5$,
and for different values of $\lambda$, close to $\lambda_c$.
In order to avoid a messy overlap, the curves have been shifted
vertically by an arbitrary value. Their asymptotic behavior
indicates good agreement with the expected DP critical scaling.
The estimates of $\lambda_c$ are reported in Table~\ref{TABQ}.
Both graphs are plotted in a doubly logarithmic scale.
}
\end{figure}
Their actual values are reported in the first line of
Table~\ref{TabSeed}. A scaling behaviour of the type described by
Eq.~\toref{spreadEQ} is expected to arise also in systems
exhibiting a first--order transition in 1+1 dimensions
\cite{HayeUnpublished}, although the relations between spreading
and stationary exponents assume a more general form than
Eq.~\toref{Hyperscaling} \cite{Hinrichsen00}. In these cases, the
critical dynamics follows from the marginal relative stability of
two coexisting phases: {\it (i)} the absorbing state itself and
{\it (ii)} a phase characterized by a suitable finite density of
active sites. For instance, an active site evolves into a droplet
of phase {\it (ii)} embedded into a sea of phase {\it (i)}. At
criticality, the boundary, having no preferential velocity,
diffuses like an unbiased random walk. A clear such example of
such a type is the one-dimensional Glauber--Ising model at zero
temperature, where the density of active sites in phase {\it (ii)}
is maximal, i.e. equal to 1. In models such as \toref{Process},
the natural existence of small inactive islands (see
Fig.~\ref{FIGDEMO}) can seemingly make the separation into two
phases questionable. However, for sufficiently small
$\sigma$-values, the long-range interactions are strong enough to
suppress the formation of large inactive islands. In these
circumstances, one would expect the long-term dynamics to be
controlled by the evolution of the boundaries. In particular, the
scaling exponents of contact processes exhibiting a first--order
phase transition in 1+1 dimensions are expected to be the same as
in the Glauber--Ising model at zero temperature, whose values are
reported in the second line of Table~\ref{TabSeed}. Notice that
$\eta=0$ means that the density remains finite at criticality.

Numerical simulations of the spreading dynamics confirm our expectations.
Independently of the results discussed in the previous section, we have first
estimated $\lambda_c$ by measuring the average number of active sites $N(t)$ for
different values of $\lambda$ and then looking for the value of the control
parameter that minimized the curvature of $N(t)$ at long times (see
Fig.~\ref{FIGN}). By performing simulations up to $t_{\mbox{max}}=3 \times 10^6$
Monte Carlo sweeps and by averaging over $10^5$ realizations, we have obtained
fairly accurate estimates, which are listed in the second row of
Table~\ref{TABQ}. The differences with the values reported in the first row
provide an indirect estimate of the magnitude of finite--size corrections.

\begin{figure}[tcb]
\includegraphics[width=125mm]{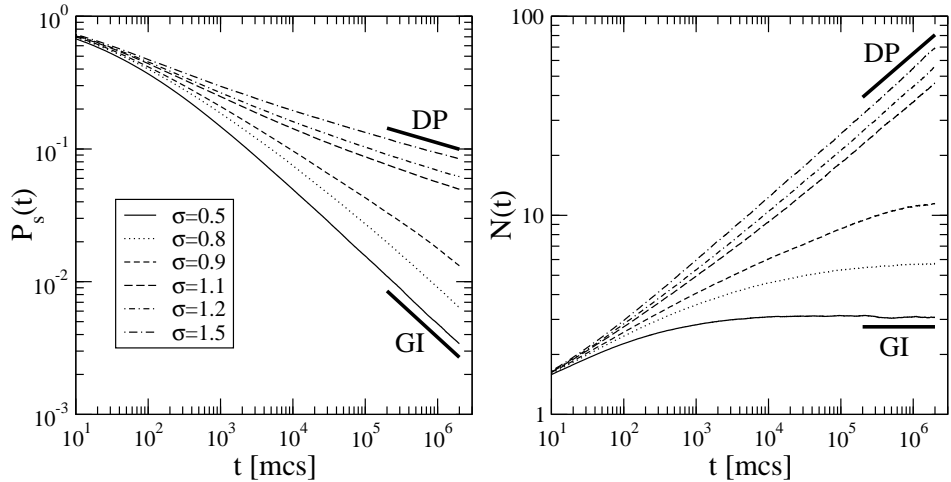}
\caption{
\label{FIGP} \small
The survival probability $P_s(t)$ (left panel) and the average number of
active sites $N(t)$ (right panel)
at criticality for $\sigma=0.5, 0.8, 0.9, 1.1, 1.2, 1.5$ (showed from bottom to
top) in a doubly logarithmic representation. The predicted asymptotic slopes of
directed percolation (DP) and Glauber-Ising (GI) are indicated as bold lines.
}
\end{figure}

We have also investigated the asymptotic behavior of $P_s(t)$ and $N(t)$
at criticality (see Fig.~\ref{FIGP}). For large times, both quantities exhibit a
power--law behavior: for $\sigma>1$ ($\sigma < 1$) the growth rates are
consistent with a DP transition (Glauber-Ising dynamics). As expected, the
closer is $\sigma$ to 1, the longer is the time needed to reach the scaling
regime.

In conclusion, our simulations indicate that the crossover from a continuous to
a discontinuous transition takes place in our model between $\sigma=0.9$ and
$\sigma=1.1$. This result is consistent with the predictions of the mean field
arguments discussed in Section 2. For the sake of completeness, we mention the
results of simulations made in the marginal case $\sigma = 1$. In such a case,
there are indications of a still first--order transition, although the
Glauber--Ising exponents are not yet recovered on the accessible time scales.

\section{Non-equilibrium wetting as a contact process with long range interactions}

In this section we investigate to what extent the behavior of non-equilibrium
wetting processes can be interpreted as a contact process with long-range
interactions of the form~\toref{Process} and whether the previous results can be
used as a criterion for distinguishing first--order from  DP--like continuous
transitions. To this end we consider two previously introduced wetting models,
studying them within the above derived framework. In practice, we numerically
estimate the effective activation rates at the boundary of detached islands
and show that they are indeed of the form~\toref{Process}. The results obtained
in the previous section can help to discern the nature of the wetting transition
in the two models.

\subsection{Restricted solid--on--solid wetting model}

\begin{figure}[tcb]
\centerline{\includegraphics[width=105mm, height=65mm]{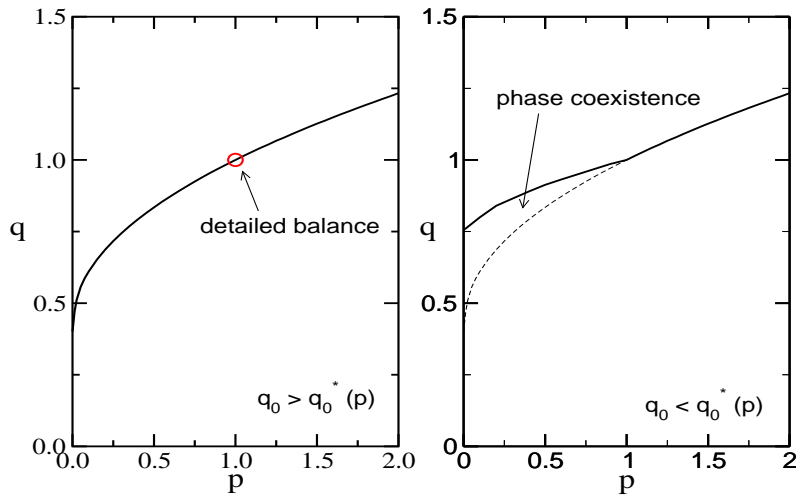}}
\caption{ \label{phasediag} \small Phase diagram of the RSOS model for
$q_0 > q_0^*(p)$ (left panel) and $q_0 < q_0^*(p)$ (right panel).
The phase--transition takes place along the full line, while
the dashed line in the right panel marks the lower border of the
phase--coexistence region (see text).}
\end{figure}

The first system we consider is a restricted solid-on-solid (RSOS)
model~\cite{Wetting1,Wetting2,Wetting3}. It is defined on a one dimensional
lattice with periodic boundary conditions: at each lattice site $i$ the height
variable $h_i$ can take any nonnegative integer value such that
$|h_i-h_{i+1}|=0, 1$. A hard attractive substrate is located at zero height
preventing $h_i$ from becoming negative. The interface evolves by random sequential
updates controlled by three real parameters $q$, $q_0$ and $p$. At each move, a
site $i$ is randomly selected and, provided the above constraints are fulfilled,
one of the following three processes is carried out~(see Ref.~\cite{Wetting3})
\begin{itemize}
\item[-] Particles are deposited with rate $q_0$ at the bottom
layer and with rate $q$ at higher layers.
\item[-] Particles
evaporate from the edges of a terrace with rate $1$.
\item[-]
Particles evaporate from the middle of a plateau with rate $p$.
\end{itemize}
The phase--diagram of the model is shown in Fig.~\ref{phasediag}. If $q_0$ is
large enough (e.g. $q_0 \approx q$), the model exhibits a line of continuous
phase--transitions, which belongs to the MN universality class. Its critical
behavior can be described by a Kardar-Parisi-Zhang (KPZ) equation \cite{KPZ} in a
potential $V(h)$ representing the interaction between the interface and the
substrate
\beq \dot{h} = D \nabla^2 h + \nu (\nabla h)^2 -{{\partial V}\over
{\partial h}}+ \zeta .
 \label{KPZeq} \eeq
Here $h\equiv h(x,t)$ indicates the height of the interface on the substrate,
while $\zeta \equiv \zeta(x,t)$ is the noise term $\delta$--correlated in
space and time, $\langle \zeta(x,t) \,\zeta(x',t')\rangle \sim \delta(x-x')\,
\delta(t-t')$. The coefficient $\nu$ of the nonlinear term is positive for $p>1$
and negative for $p<1$. Detailed balance holds only for $p=1$. This special case
can be solved exactly \cite{foot3}: its critical properties are described by
an Edwards-Wilkinson (EW) equation~\cite{EW} ($\nu=0$) equipped with the
interface--substrate interaction potential.

Decreasing $q_0$ amounts to increasing the attractive interaction
between interface and substrate. It has been observed that, below
a certain threshold, the continuous transition may turn into a
discontinuous one~\cite{Wetting2}. For instance, at $p=1$ it was
shown analytically that for $q_0<2/3$ the transition becomes
first--order, while the transition point remains located at $q=1$.
Numerical simulations provide very clear evidence that this
scenario extends to the non--equilibrium case $p>1$: the
phase--transition line is still independent of $q_0$ (see
Fig.~\ref{phasediag}), while for $q_0$ smaller than a threshold
value $q_0^*(p)$, the transition becomes first--order. In this
case, the interface dynamics is quite different from the one
described by our model. In fact, direct inspection of the
interface dynamics in the active phase close to criticality shows
that there is a non--negligible probability for the interface to
return in contact with the substrate not only at the domain
boundaries of inactive domains, but also inside these domains. In
DP jargon, this amounts to saying that an inactive site may become
active even without being in contact with an active site. This
excludes any direct relation with the model of a generalized
contact process with long--range interactions introduced in
Section 2.

Conversely, for $0<p<1$, the dynamics appears to be strongly related to that of
our model. For sufficiently small $q_0$ (i.e. $q_0$ smaller then a threshold
$q_0^*(p)$), a region in the $(p,q)$ plane arises,
where the pinned and the unbound phases coexist (see Fig.~\ref{phasediag}b). In
this region, an unbound interface moves away from the substrate and never binds
back. On the other hand, a bound interface remains bound for macroscopically
long times and, in the thermodynamic limit, will never detach from the
substrate. Only if the growth rate $q$ is increased beyond a new critical value,
a depinning transition takes place at the upper border line of the
phase--coexistence region (see Fig.~\ref{phasediag}b). It is the nature of this
transition that we want to investigate here.

As explained in~\cite{Wetting2,Wetting3}, the stability of the pinned (active)
phase in the coexistence region is ensured by the negative sign of the
nonlinear coefficient $\nu$: when a large detached island forms, it grows
quickly, acquiring a triangular shape with a given slope and eventually shrinks
from the edges with constant velocity. Because of the triangular shape of the
interface, the probability of returning to the substrate at some point far from
the edges of the island is exponentially suppressed. Therefore, it is
reasonable to conjecture that for $p<1$ the model belongs to the class of
contact processes.

Numerical simulations of the RSOS model dynamics (not shown here) seem to
indicate that also in this case the depinning transition may become
discontinuous for $q_0<q_0^*(p)$. For $p=0$ and, correspondingly,
$q_0<q_0^*(0) \approx 0.399$, we know that it is of DP type \cite{Wetting2}.
On the other hand, the DP scaling regime becomes transparent only after a
transient time that, for $q_0 = 0.35$ is on the order of $10^4$ units.
Although the crossover becomes practically unobservable for yet smaller
$q_0$-values, there are compelling reasons to believe that this regime is
eventually attained~\cite{Wetting2}.

In order to shed some light about the order of the transition for $0<p<1$, we
have tested whether effective long-range effects spontaneously emerge as a
result of the RSOS microscopic dynamics. In practice, we have measured the
effective activation rate $\bar{\lambda}(\ell) = N_a(\ell)/ N_b(\ell)$ at the
border of depinned islands of size $\ell$ ($N_b(\ell)$ is the number of times a
depinned site at the border of an inactive island of size $\ell$ has been
selected in the stationary regime and $N_a(\ell)$ is the number of times the
selected site is immediately pinned). Data has been obtained by averaging over
time, space and different realizations (typically, 100) for large lattices
($L=10^5$) and close to the transition line.
\begin{figure}[tcb]
\centerline{\includegraphics[width=8cm]{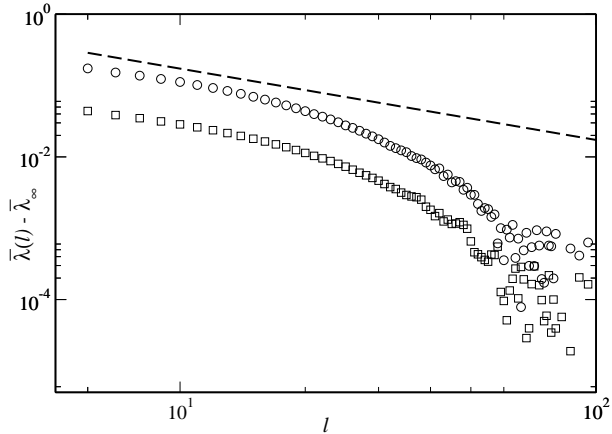}} \caption{
\label{PSmall} \small RSOS model - decay of the activation rate (see
text) as a function of size of depinned islands. Numerical
simulation have been performed close to the special point $p=0$.
Circles refer to $T_0 = (p=0, q_0=0.2, q=0.755)$, while squares
to $T_1=(p=0.01, q_0=0.2, q=0.76)$. The graph is plotted in a
doubly logarithmic scale. For the sake of clarity the data are
shifted vertically by an arbitrary value.}
\end{figure}
\begin{figure}[tcb]
\centerline{\includegraphics[width=8cm]{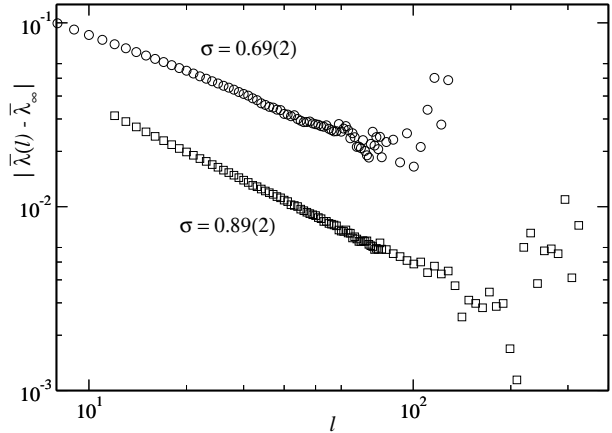}} \caption{
\label{P02} \small RSOS model - decay of the activation rate (see text)
as a function of size of depinned islands. Numerical simulation
have been performed at the critical points
$T_2=(p=0, q_0=0.4, q=0.70)$ (squares)
and $T_3=(p=0.2, q_0=0.3, q=0.745)$ (circles).
The graph is plotted in a doubly logarithmic scale.
Data at $p=0.3$ show a power law decay with an exponent $\sigma =
0.69(2)$, while $q_0=0.4$ data decays with an exponent $\sigma =
0.89(2)$. }
\end{figure}

We have first analyzed the nature of the transition close to $p=0$. In
Fig.~\ref{PSmall} we display the results obtained at the transition points
$T_0=(p=0, q_0=0.2, q=0.755)$ and $T_1=(p=0.01, q_0=0.2, q=0.76)$. The activation
rate $\bar{\lambda}(\ell)$ is reported after subtracting its estimated
asymptotic value $\bar{\lambda}_{\infty} = \lim_{\ell \to \infty}
\bar{\lambda}(\ell)$. In both cases, it is found that it converges towards
$\bar{\lambda}_{\infty}$ faster than $1/\ell$, although no precise estimate of
the scaling rate can be obtained. Altogether, these results suggest that DP
critical properties persist also for small values of $p$ and $q_0<q_0^*(p)$.
On the other hand, for $p=0.2$ and $q_0 < q_0^*(0.2)=0.515\ldots$ our findings
are suggestive of a first--order phase transition (and thus, in agreement with
previous findings \cite{Wetting2}). In fact, from Fig.~\ref{P02}, we see that in
both points $T_2=(p=0.2, q_0=0.4, q=0.70)$, and $T_3=(p=0.2, q_0=0.3, q=0.745)$
the activation rate
is found to converge slower than $1/\ell$, although the actual value of the
exponent $\sigma < 1$ appears to depend on the parameter. According to the
criterion introduced in Sec. 2, these results are therefore compatible with a
discontinuous phase--transition.

As a result, we can attribute the seemingly first-order nature of
the transition to the existence of effective long-range
interactions. Nevertheless, it remains to be proved whether the
slow dependence of the activation rate on the window size is just
a finite-size effect or holds fot arbitrarily large distances.
Here below we present an argument supporting the former
hypothesis.

In order to clarify this point we move progressively away from the
equilibrium case. For $p=1$, it is known that the transition is
discontinuous~\cite{Wetting3} (for small enough $q_0$) and that
the dynamics of the free interface is asymptotically described by
the EW equation. When $p$ is lowered below 1, the only relevant
difference that is expected to occur is a crossover in the free
interface dynamics from and EW to a KPZ regime above some critical
length $\ell_c$ \cite{Barabasi, SKJJB92}. It is thus natural to
conjecture that as long as the dynamics of the bound interface
remains insensitive to such differences, the scenario should not
change. Here below we argue that this occurs for system sizes
smaller than some length $L_c$ that can be extremely large.

The best way to characterize the above mentioned crossover is by
monitoring the width $ w(t) = (\langle (h - \langle h \rangle)^2
\rangle)^{1/2}$ of an initially flat (and free) KPZ interface. In
fact, after an initial growth as $t^{1/4}$ (in agreement with EW
equation), at some time $t_c$, $w(t)$ crosses over towards a
behavior of the type $t^{1/3}$ and eventually saturates because of
the finite length of the system. In other words, for times smaller
than $t_c$ the interface behaves in the same way as in an
equilibrium regime. Since the stationary profile of an interface
is a diffusive random walk, one can safely assume that \beq
w_\infty = \lim_{t \to\infty} w(t) = k \sqrt{L} \label{satwidth}
\eeq where $L$ is the interface length. By extending the above
relation to finite times, one can interpret it as the definition
of the effective scale $L(t)$ that is resolved at time $t$. For
instance, $\ell_c = (w(t_c)/k)^2$ is the minimal length of a free
interface that allows observing a crossover to the KPZ scaling
behavior.

Within the context of a bounded interface, this implies that deviations from
equilibrium are observable only in those depinned islands of length
$\ell > \ell_c$. Accordingly, the problem of determining the minimal length to
observe deviations from the equilibrium scenario amounts to estimating the
probability for a suitably large depinned islands to arise. At equilibrium, the
theoretical analysis developed in \cite{Wetting3} has revealed that when the
transition is discontinuous (the scenario we are interested in), the
interface is exponentially localized at the substrate, i.e. the probability to
find large values of the height $h$ scales as $P(h) \simeq \exp(-h/h_0)$.
It is quite plausible to assume, and we have indeed numerically verified, that as long as
deviations from equilibrium are not detectable, the exponential decay survives
also for $p<1$. Since we have also seen that depinned islands have an
approximately triangular shape, this means that also island lengths are
exponentially distributed $P(\ell) \simeq \exp(-\ell/\ell_0)$, where $\ell_0$ is
proportional to $h_0$, the proportionality constant being related to the slope
of such islands.

As a result, the probability that at least a given island reaches the
size $\ell_c$ is proportional to $\exp(-\ell_c/\ell_0)$. In a
large but finite system of size $L$, this may happen independently at
different places. Hence the first large island would appear in a typical time
$\tau \simeq \ell_c/L\,\exp(\ell_c/\ell_0)$.
Accordingly, the minimum system size guaranteeing that such islands
are observed with nonnegligible probability and dominate the wetting dynamics
is
\beq
L_c = \ell_c \, \exp(\ell_c/\ell_0)\,.
\label{Lc}
\eeq
Therefore, as long as the
(effective) size remains smaller than $L_c$, a seemingly first-order transition
is observed. Beyond $L_c$, in the fully nonlinear regime, several theoretical
and numerical studies of different models \cite{Ginelli03, Munoz} suggest that a DP behavior
sets in. As a result, we expect in particular that the distribution of depinned
islands crosses over from an exponential to a power-law distribution. However,
given the exponential dependence of $L_c$ on $\ell_c$, it may happen that the
crossover length is so large that the asymptotic regime is practically
unobservable. Note also that $L_c$ diverges exponentially
as $p \to 1$, thus approaching the equilibrium point.

Numerical simulations performed at the transition point
$T_1 =(p=0.01, q=0.76, q_0=0.2)$ indicate that $k=0.179(1)$ and
$w(t_c)=2.9(2)$, yielding the estimate $\ell_c \approx 260$.
On the other hand, direct computations of the island sizes indicate that
$\ell_0 \sim {\cal O}(10^2)$, yielding $L_c \approx 3.5 \times 10^3$ and
altogether confirming that the crossover towards
DP can be observed, as we actually do. Moreover, at the transition point
$T_2 = (p=0.2, q=0.70 , q_0=0.4)$, we find $k=0.204(1)$ and $w(t_c)=5.7(2)$,
while $\ell_0$ is almost unchanged,
yielding the estimates $l_c \approx 780$ and $L_c \approx 2 \times 10^6$.
Consistently, no indication of the crossover has been observed up to times on
the order ${\cal O}(10^6)$ and system sizes of length ${\cal O}(10^5)$.

\subsection{Single--Step--with--Wall model}

\begin{figure}[tcb]
\centering
\includegraphics*[width=8cm, angle=0]{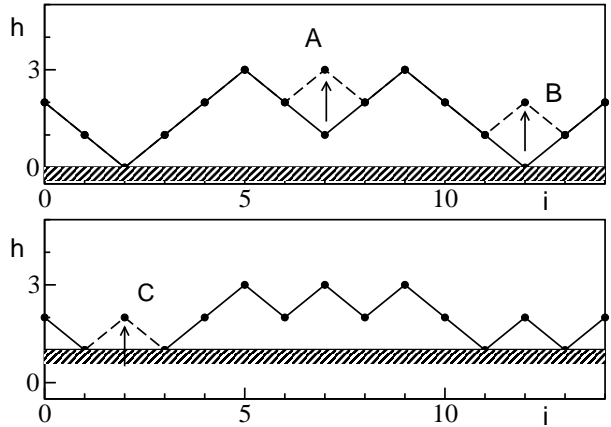}
\caption{\small
Updating rule of the SSW model. The full line represents
the interface, while the shaded area represents the wall. Dashed
segments indicate interface growth occurring in randomly chosen
local minima (see A and B in upper panel) and in all sites located
below the wall after it has been shifted upwards by one unit (see
C in lower panel). } \label{SSWsketch}
\end{figure}

The second model we have tested is the so-called ``single-step-with-wall''
(SSW) model~\cite{Ginelli03}, a variant of the well known Single--Step model
introduced in Refs.~\cite{MRSB,KS}. Here, the growing interface is described by a
set of integer heights $h_i$ at site $i$ of a one-dimensional lattice of length
$L$ with periodic boundary conditions. In this model, the ``continuity''
restriction $|h_i-h_{i+1}|=1$ plays the role the RSOS constraint. An
upward--moving wall is located at some integer height $h_w(t)$, below the
interface. It moves with velocity $v_w$, thus pushing the interface which
cannot be overtaken by the wall. Moreover, in analogy with the RSOS model, the
interface is also attracted by the wall. The model evolves by random-sequential
dynamics, i.e., at each time step $dt=1/L$, a site $i$ is chosen at random. If
the interface has a local minimum at site $i$ (namely $h_i < h_{i\pm 1}$), the
height $h_i$ is increased by two units with probability $1$ if $h_i > h_w$, or
with probability $(1-q)$ whenever the interface is pinned to the wall
($h_i = h_w$). Since $0< q< 1$, an effective attractive force is introduced
between the wall and the interface.
After $n_w=L/v_w$ time steps, the wall is moved upward by one unit, while
the height of all interfacial sites that would be overtaken by the
wall is increased by two units (e.g., see Fig. \ref{SSWsketch}).
On the basis of these microscopic update rules, one can easily
infer that the velocity of the free interface is $1/2$ in the thermodynamic
limit.

\begin{figure}[tcb]
\centerline{\includegraphics[width=8cm]{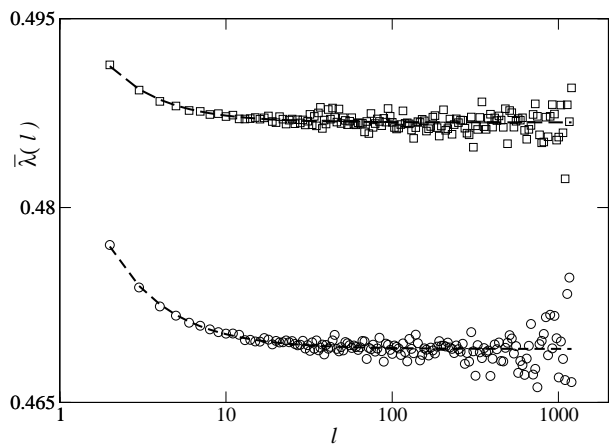}} \caption{
\label{SSWFig} \small
SSW model - Double logarithmic plot of the activation
rate $\bar{\lambda}(\ell)$ as a function of the size
$\ell$ of depinned islands. Numerical simulations have been
performed at criticality: circles corresponds to $q=0.7,\quad
v_w=0.4297$ while squares to $q=0.8,\quad v_w=0.3489$. Dashed
lines mark our best fits with Eq. \toref{Process},
rendering respectively the estimates $\sigma = 1.20(5)$ and
$\sigma = 1.30(5)$. }
\end{figure}

The phase--diagram of the SSW model is controlled by two parameters: $v_w-1/2$,
i.e. the relative velocity of the wall to the free-interface, and $q$, i.e.
the ``stickiness'' of the wall. By decreasing $v_w$, a phase--transition
from a pinned to a depinned phase is observed. Numerical analysis and
analytical arguments \cite{Ginelli03} show that when $q<q^*=0.444 \ldots$ such a
transition takes place at $v_w^c=1/2$, i.e. when the relative velocity of the
wall with respect to the free-interface changes its sign. The transition in
this part of the phase--diagram is continuous and belongs to the MN universality
class. For $q>q^*$ the effective attractive force binds the interface to the
wall and a continuous DP  phase--transition takes place at a critical value
$v_w^c(q) < 1/2$.

In order to estimate the role of long range interactions in the critical
dynamics of the SSW model, we have measured also in this case the effective
activation rate $\bar{\lambda}(\ell)$. Since the time interval $1/v_w$ between
two consecutive wall moves can be regarded as the natural time scale of the
SSW model, we measured the pinning rate at time $t_m=m/v_w$ ($m=1, 2, \ldots$)
for an initially depinned site located at the border of a depinned island of
size $\ell$ at time $t_{m-1}$. Averages have been taken over time, space and
different ensemble realizations (typically 100) for large enough systems
($L=10^5$) close to the DP critical line, at $q=0.7$ and $q=0.8$. Our results
are shown in Fig.~\ref{SSWFig}, where numerical data has been fitted with
$\bar{\lambda}(\ell)=\lambda(1+a/\ell^\sigma)$. There is evidence that
$\bar{\lambda}(\ell)$ decays to a constant faster than $1/\ell$ with exponents
$\sigma = 1.20(5)$ (for $q=0.7$) and $\sigma = 1.30(5)$ ($q=0.8$). According to
our predictions based on the behavior of the generalized contact process, this
implies a continuous DP transition, which indeed has been observed in
Ref.~\cite{Ginelli03}.

Unlike the RSOS model discussed in the previous section, here in the SSW model
the DP behavior is observed at small length scales, and is thus accessible in
numerical studies of finite systems. The reason is that the SSW model is
designed in a way that the nonlinearity is maximal \cite{Barabasi}, and hence
the crossover length $\ell_c$ is fixed and of order $1$. On the other hand in
the RSOS model the nonlinearity depends on the growth parameter $p$, with a
diverging crossover length as $p \to 1$.

\section{Conclusions}

In this work we examined a possible connection between wetting phenomena and
contact processes. Inspired by the puzzling richness of the phase--diagrams
found in various models, we introduced a generalized contact process with the
goal of capturing both the DP and first-order transitions observed in
one-dimensional non-equilibrium wetting transitions occurring at non-zero
interface velocity. The element of novelty distinguishing our model from
standard contact processes consists in an algebraic dependence of the
activation rate on the length $\ell$ of the depinned island containing the
inactive site $\bar \lambda(\ell) \simeq \lambda(1+a/\ell^\sigma$).

A mean--field analysis predicts that when $\sigma>1$, the model exhibits a
continuous phase--transition characterized by DP critical exponents. In other
words, the algebraic decay of the interactions is not so long-range as to alter
DP critical properties. Conversely, for $0<\sigma<1$ the phase--transition
turns to a first--order one. Numerical simulations confirm the mean--field
predictions, which are found to provide also an accurate estimate for the
critical value $\sigma_c$, separating the two different regimes:
$\sigma_c = 1.0 \pm 0.1$, from numerics. By directly measuring the effective
activation rate $\bar{\lambda}(\ell)$ in non-equilibrium wetting processes, one
can use the estimate of the exponent $\sigma$ as a practical tool for probing
the nature of a wetting transition.

\begin{figure}[tcb]
\centerline{\includegraphics[width=9cm]{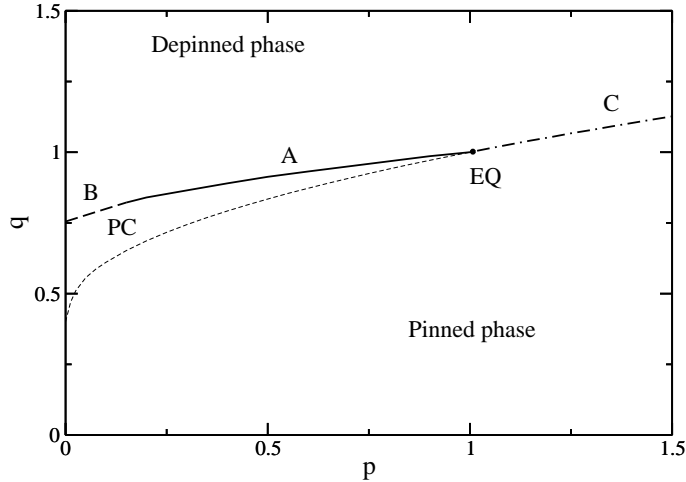}} \caption{
\label{Sketch} \small Qualitative ``finite--size'' phase--diagram of the
RSOS model in the case $q_0<q_0^*(p)$. We expect the entire
transition line for $p<1$ to belong to the DP universality class.
The full line (line A) on the left of the equilibrium point (EQ)
marks a region in which the transient first--order behavior is
the only numerically accessible one, with a crossover
lenght $L_c$ to DP behavior which diverges as $p \to 1$.
On the other hand, also systems of moderate size
can exhibit a DP critical behavior over the
dashed line (B). The dot dashed transition line (C) on the right
of the equilibrium point is genuinely first--order (see text).
Finally, the light dashed line where the free interface velocity
changes sign marks the lower boundary of the phase--coexistence
region (PC).}
\end{figure}

To apply the insight gained from the DP process to wetting phenomena, we
considered in this paper two wetting models, RSOS and SSW (see Section 4).
Previous numerical studies of the latter model \cite{Ginelli03} indicate that for
a sufficiently strong attractive force of the substrate (wall), the wetting
transition is DP. Our criterion confirms these results, since the effective
activation rate is indeed found to converge faster than $1/\ell$ to its
asymptotic value. The phase--diagram of the RSOS model on the other hand is
known to be more complicated, since it contains both first order and
continuous wetting transitions\cite{Wetting2}. Our analysis suggests that for
$p<1$ the entire phase--transition line located at the upper border of
the coexistence region should asymptotically belong to the DP
universality class. However the finite--size behavior depends on two length
scales: ({\it i}) the crossover $\ell_c$ between EW and KPZ roughening behavior;
({\it ii}) the effective size $\ell_0$ of depinned islands. As long as the
system size is smaller then $L_c = \ell_c \, \exp(\ell_c/\ell_0)$ and
islands of size larger than $\ell_c$ are not significatively
generated, numerical
simulations can only provide evidence of a first-order transition.
A qualitative  phase--diagram for the RSOS model in the case $q_0 < q_0^*(p)$ is
sketched in Fig.~\ref{Sketch}.

It is interesting to compare our results with those recently obtained for the
KPZ equation with an attracting hard core potential~\cite{Munoz}:
$\dot{h} = D \nabla^2 h + \nu (\nabla h)^2 - V'(h) + \zeta$. Although the
authors do not exclude the possibility that the entire transition line belongs
to the DP universality class, their numerical analysis revealed both a first
order and a DP wetting transition. In particular, DP critical behavior has
been observed upon decreasing the attractive force and reducing by a factor
ten {\it both} the diffusion ($D$) and the nonlinear ($\nu$) term with respect to
the parameter values corresponding to a first--order phase--transition. While
decreasing the attractive force is obviously expected to increase the ``cut-off'' scale
$\ell_0$, the latter change decreases the length scale $\ell_c$ by the same
factor ten. The crossover length from EW to KPZ roughening is in fact known to
scale as $D^3 / (\nu^2 \Gamma)$ \cite{SKJJB92}, where $\Gamma$ is the amplitude of
the Gaussian white noise $\zeta$. This indeed suggests that the DP
(first--order) behavior has been observed in a parameter range where the
crossover scale $L_c$ (as defined by Eq. \toref{Lc}) is numerically accessible (unaccessible),
as suggested in the present study.
This work helps to better understand the part of the wetting phase diagram
characterized by a negative coefficient of the KPZ nonlinearity, which is
precisely the part related with the complete synchronization transition 
(in the case of an MN phase transition, this relation can be made
explicit by use of the Cole-Hopf transformation \cite{PikovPoliti}).
In particular our analysis indicates that the dynamical details of the
system may induce a seemingly first order phase transition which lasts
over exponentially long time and space scales, effectively suppressing 
DP critical properties. \\

Finally, we wish to comment about the transition line for $p>1$ (where the
coefficient of the KPZ nonlinearity is positive). For
$q_0<q_0^*(p)$, the transition is known to be first--order, but it
occurs when the interface velocity changes sign~\cite{Wetting2},
i.e. there is no region of phase--coexistence. Accordingly,
depinned islands are no longer characterized by a triangular shape
and interface fluctuations may easily give rise to the pinning of
inactive sites far away from the active ones; in other words, the
analogy with contact process is seemingly lost. It would be
interesting to investigate whether the inclusion of some sort of
``spontaneous'' nucleation of active sites can eventually account
for the scenario observed for $p>1$. A further open problem is the
crossover from the DP to the MN universality class, which takes
place in both RSOS and SSW models. The study of this crossover
would require the inclusion of nucleation of active sites in the
interior of inactive domains. An appropriate generalization of the DP
model including such processes could yield useful
insight onto the crossover phenomena taking place in non-equilibrium
wetting.

{\bf Acknowledgements} We wish to thank A. Torcini for many
stimulating and fruitful discussions. The support of the Israeli
Science Foundation (ISF), the Minerva Foundation and the Albert
Einstein Center for Theoretical Physics is gratefully
acknowledged.



\begin{thebibliography}{98}
\bibitem{foot1}
The same power-law
behavior for $P(\ell)$
can be obtained by a general theorem on fractals, which relates
the size distribution of inactive islands to fractal dimensions
\cite{AharonyMandelbrot}. A scaling argument for the space--like fractal
dimension of DP can be found in ref.~\cite{Gras89}.
\bibitem{foot2}
        It should be noted that the density of active sites $\rho(t)$ and
        the survival probability $P_s(t)$ decay with the same critical
        exponent $\theta=\delta$
        due to a special symmetry of DP \cite{Hinrichsen00}.
\bibitem{foot3}
For completness, we mention that the case $p=0$ is another special
point of the phase--diagram, since here the dynamical rule does not
allow for particle evaporation in the middle of depinned portion
of the interface. Numerical analysis and the comparison with an
analogous unrestricted model \cite{Alon} strongly suggests that it
belongs to the DP universality class.
\end{thebibliography}
\end{document}